# Predicting the meal macronutrient composition from continuous glucose monitors

Zepeng Huo[1], Bobak J. Mortazavi[1], Theodora Chaspari[1], Nicolaas Deutz[2], Laura Ruebush[2], Ricardo Gutierrez-Osuna[1]
[1]Department of Computer Science and Engineering
[2]Center for Translational Research in Aging and Longevity
Texas A&M University, College Station, Texas, USA
{guangzhou92, bobakm, chaspari, nep.deutz}@tamu.edu, le.rubush@ctral.org, rgutier@cse.tamu.edu

*Abstract*— Sustained high levels of blood glucose in type 2 diabetes (T2DM) can have disastrous long-term health consequences. An essential component of clinical interventions for T2DM is monitoring dietary intake to keep plasma glucose levels within an acceptable range. Yet, current techniques to monitor food intake are time intensive and error prone. To address this issue, we are developing techniques to automatically food intake and the composition of those foods using continuous glucose monitors (CGMs). This article presents the results of a clinical study in which participants consumed nine standardized meals with known macronutrients amounts (carbohydrate, protein, and fat) while wearing a CGM. We built a multitask neural network to estimate the macronutrient composition from the CGM signal, and compared it against a baseline linear regression. The best prediction result comes from our proposed neural network, trained with subject-dependent data, as measured by root mean squared relative error and correlation coefficient. These findings suggest that it is possible to estimate macronutrient composition from CGM signals, opening the possibility to develop automatic techniques to track food intake.

## I. INTRODUCTION

One hundred and twenty million Americans are glucose intolerant because the body does not respond sufficiently to insulin. The mild form of this intolerance is known as pre-diabetes, whereas the severe form is known as type 2 diabetes mellitus (T2DM). High plasma/blood levels of glucose in the morning and after meals can have disastrous long-term health consequences, including cardiovascular diseases (the main cause of death in the developed world), retinopathy (leading to blindness), peripheral neuropathy (leading to limb amputations), and nephropathy (kidney damage). An essential component of clinical interventions for diabetes is monitoring dietary intake to keep plasma glucose levels within an acceptable range after a meal. However, conventional methods for diet tracking rely on human recall and/or manual input, which are inaccurate and burdensome –regardless of the medium used for logging (i.e., paper or electronic).

A unique and unexplored opportunity to solve this problem has emerged with the advent of continuous glucose monitors (CGMs). A CGM consists of a small electrode inserted under the skin (subcutaneously), and a transmitter that sends the information to a monitoring device. What makes CGMs particularly appealing is the fact that the plasma glucose response to a meal depends on its macronutrient composition (i.e., carbohydrates, proteins, fats); as an example, combining fat and protein with carbohydrates generally leads to smaller increases and slower decreases of glucose concentrations [1]. This suggests that the shape of the glucose response to a meal can be used to recover the macronutrient composition of the meal, and therefore be used to log food intake automatically.

To test this hypothesis, we have conducted a clinical study where pre-diabetic participants were asked to consume a set of standardized meals (i.e., of known carbohydrate, protein and fat). Then, we measured their blood glucose response after each meal using a CGM. Finally, we built a neural network model to estimate the macronutrient composition from the CGM signals. Our results show that subject-dependent models are more accurate than subject-independent models, especially if models account for interaction effects among macronutrients.

## II. RELATED WORK

### A. Continuous glucose monitors

To manage diabetes, patients must take frequent measures of blood glucose by pricking the finger with a lancet –a painful procedure that must be repeated several times per day. CGMs, however, typically measure glucose automatically every 5 to 15 minutes, which dramatically reduces burdens to the patient and, with 288 glucose measurements per day, make it possible to track the effect of each meal almost automatically. CGMs have gained acceptance to manage type-1 diabetes but have yet to make an impact in T2DM [2], by far the most predominant of the two diseases (90%). An often-cited limitation of CGMs is the need to calibrate them regularly with a finger-prick blood measurement, sometimes up to 3-4 times per day [3]. Fortunately, this is no longer the case for the newer generation of CGMs, which can be worn uninterruptedly for up to 10-14 days without recalibration. A second limitation of CGMs is their cost, which can be in the $800 range for a starter kit and $70 per disposable sensor. Fortunately, the new generation of "flash" CGMs are relatively inexpensive; as an example, the Abbott Freestyle Libre costs $65 for a reader and $60 per disposable sensor, and many of these CGMs are now covered by Medicare.

### B. Diet monitoring technology

Several recent surveys have examined solutions to diet monitoring technology, from smart utensils [4] to wearable sensors to track eating habits [5] and computer vision techniques to visually interpret diet intake [6]. As an example, wearable sensors have been used to log food intake [7] by detecting specific gestures used to eat the food. Acoustic methods with a variety of sensors placed on the body have also been used to log food intake [8-10], though these methods also focus on detecting swallows and coarse categories of food intake. While eating episodes can be detected accurately [11-13], wearable solutions to estimate nutritional content are not available.

TABLE I. COMPOSITION OF MEALS IN THE STUDY. GREEN, YELLOW AND RED INDICATE LOW, MEDIUM AND HIGH VALUES, RESPECTIVELY.

| Meal | Carbohydrate (g) | Protein (g) | Fat (ml) |
|---|---|---|---|
| Meal1 | 52.25 | 15 | 13 |
| Meal2 | 94.75 | 30 | 26 |
| Meal3 | 179.75 | 60 | 52 |
| Meal4 | 52.25 | 30 | 26 |
| Meal5 | 179.75 | 30 | 26 |
| Meal6 | 94.75 | 60 | 26 |
| Meal7 | 94.75 | 15 | 26 |
| Meal8 | 94.75 | 30 | 52 |
| Meal9 | 94.75 | 30 | 13 |

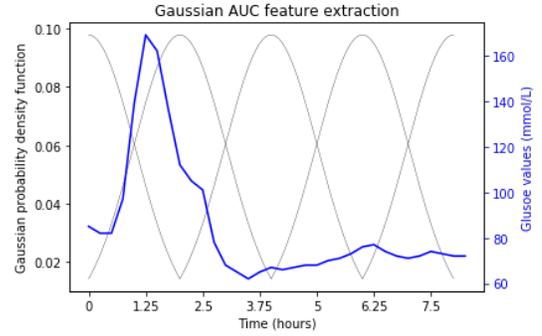

Fig. 1. Post-prandial glucose response (blue) and family of Gaussian kernels used to compute the AUC at various time points.

*C. Modeling glucose responses*

Differences in macronutrient intake lead to different postprandial glucose responses. Previous studies have found that simple carbohydrates result in steep glucose spikes, while proteins and fats yield smoother fluctuations [14, 15]. A study examining 38 foods with isoenergetic portions found that protein-rich foods produced the highest insulin secretion per gram of carbohydrate, followed by bakery products, snack foods, fruits, carbohydrate-rich foods, and cereals [16]. Significant differences of insulin and glucose responses for the same type of food have also been found among subjects, suggesting the importance of designing models that account for individual differences. Based on these findings, there is a line of work that has performed personalized meal prediction using CGM recordings, daily habits, and person profiling [17, 18]. In contrast with these studies, which focus on predicting CGM responses to different types of meals, our goal is the reverse: predicting meal composition from CGM signals.

III. METHODS

*A. Data collection*

We recruited seven subjects ages 60-85 years and Body Mass Index in the range of 25-35. Each subject participated in 9 study days in which they consumed a predefined meal in a randomized design. Each study day lasted approximately 8 hours and the procedures on the study days were identical, with the only change being the macronutrient composition of the meal taken, varying an average diet meal with low and high values of carbohydrates, proteins, and fats. Subjects were asked to fast prior to the 9 study days, so that the first blood glucose reading would be their fasting glucose. The CGM was placed on the first study day and replaced every 2 weeks. After taking a baseline blood sample the morning of a study visit, a predefined meal was consumed. The blood samples served to validate the CGM reading accuracy. The composition of the nine meals is shown in **Error! Reference source not found.** This study was approved by the Texas A&M Institutional Review Board (IRB #2017-0886).

*B. Feature Extraction*

To capture the characteristic shape of the PPGR from the consumed meals, we computed the area under the curve (AUC) at five distinct time points, as illustrated in Fig. 1. These five time points captured the fasting glucose level, the rise in the sensor response, intermediate values when returning to fasting levels, the drop in glucose, and the final glucose levels.

To ensure that the 95% confidence interval of the Gaussian kernel aligned with these time intervals, we set the variance of the Gaussian kernel to σ= n/1.96 where n is the number of sensor readings in that time interval (the CGM records data every 15 minutes). Once the variance was fixed, we varied the location of each kernel (i.e., its mean) to cover the entire 8-hour PPGR.

*C. Model development*

We performed two types of validation: leave-one-subject-out (LOSO) and leave-one-meal-out (LOMO). In LOSO, we built a model based upon data from six subjects and their PPGRs from all meals, and tested on the seventh individual's meals. In LOMO, we used data from eight meals for each participant and tested on the ninth meal and repeated this process across all nine meals and each participant. This meant that, for LOMO, each participant had her/his own model. This allowed us to test whether subject-dependent models or subject-independent models provided better estimates.

We evaluated two techniques to compare regression of macronutrient composition from the PPGR, a baseline linear technique (least squares regression), and a non-linear technique (multitask neural network). The linear regression technique involved developing three separate regression models to estimate carbohydrates, proteins, and fats. The multitask neural network also produced three estimates, but based upon a single model. The input to both models (i.e., linear regression and multitask neural network) were the five extracted AUC values.

The multitask neural network contained two layers: a fully-connected shared layer, designed to learn the shared impact among the three macronutrients, and a task-specific layer, designed to estimate each macronutrient. The neural network was trained to minimize the Huber loss, which combines L1 and L2 losses, using L1 when error is large and L2 otherwise.

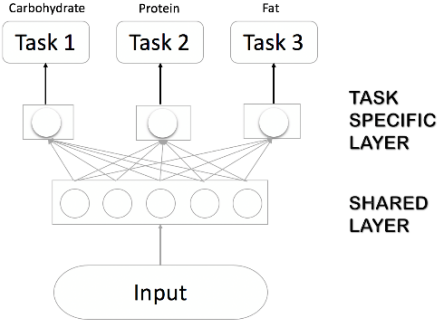

Fig. 2. Multitask neural network for estimating macronutrient composition from PPGRs.

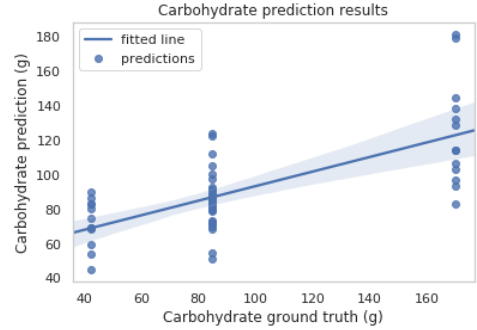

Fig. 3. Results of carbohydrate estimation, correlation, and 95% confidence interval in a leave-one-meal-out setting.

The activation functions consisted of Rectified Linear Units (ReLU) for the shared layer, and linear units for the final, task-specific layer. We grid-searched the hyperparameters for the neural network. For the number of neurons in the shared layer, we searched the interval [2,5], because of a limited amount of input data features, to avoid overfitting. We grid searched the learning rate from the set [0.0005, 0.001, 0.005]. We did a nested cross-validation on the training data to find the optimal hyperparameters. We trained for 1,000 epochs. We evaluated our grid-searched results using both the Pearson correlation between the estimated macronutrient composition and ground truth. We evaluated the final model with the Pearson correlation as well as the root mean square relative error (RMSRE) for the estimates. The root mean square relative error is calculated as:

$$RMSRE = \sqrt{\frac{1}{s}\sum \frac{(y-\hat{y})^2}{y^2}} \quad (1)$$

where s is the number of samples, $y$ the true quantity and $\hat{y}$ the estimate. We used the RMSRE because of the different quantities of carbohydrates, protein, and fat.

## IV. RESULTS

We conducted the experiments on data collected from the seven participants for carbohydrates (CHO), protein, and fat. We computed the Pearson correlation by first gathering all estimated data into a single plot, and then calculated a single correlation. Fig. 3 shows the correlation obtained for carbohydrates using the multitask neural network and the LOMO model.

First, we conducted the LOSO cross-validation experiments, in order to evaluate model differences in the macronutrient composition. The left column of Tables II and III show the LOSO correlation and RMSRE results. The correlation results also indicate the statistical significance of the correlation. Both the linear regression and multitask neural network have similarly significant correlations on carbohydrates, at 0.32 and 0.31 respectively. The error between the models is similar.

Tables II and III show the results of the LOMO cross-validation experiments in the right column. The linear regression did not see improvements. In fact, the error actually increased, indicating the personal models had a difficult time estimating new meals. The multitask neural network model, however, saw significance improvements in correlation and error. The correlation for carbohydrates is 0.69 for LOMO versus 0.31 for LOSO. Both protein and fat have a statist ically significant correlation for the multitask neural network, at 0.23 and 0.48 respectively. The RMSRE also decreased.

## V. DISCUSSION

We have proposed a multitask neural network that can be used to predict the macronutrient composition of meals from CGM data. The network consists of a shared layer that learns information common to the three macronutrients, and a task-specific layer that is customized to each macronutrient. We evaluated the network using two cross-validation procedures: leave one meal out (subject dependent) and leave one subject out (subject independent).

The subject-dependent results achieved higher predictive accuracy than the subject-independent model, despite the fact that it had fewer examples for training, suggesting that the subject-dependent factors (e.g., metabolism, age, gender, health condition, and lean body mass) require personalized modeling. This is further demonstrated by the similar correlations and errors found in the LOSO cross-validation experiments. The correlations and errors in carbohydrate estimates demonstrate that subject-dependent factors play a larger role in estimation from CGM signals than features extracted from the signal itself.

The multitask neural network achieved higher predictive accuracy in the LOMO cross-validation experiments. This would suggest that the subject-dependent models, which are not accountable to subject-to-subject variations in subject-dependent factors, can more accurately model estimates of macronutrient composition from CGM signals. This would suggest that nonlinearities in food metabolism also need to be taken into consideration.

A number of future directions for this work are possible. More subject-dependent factors should be incorporated in order to better understand the impact these factors have on the post-prandial glucose response. Additionally, we can explore the quantity of known meals needed to rapidly train subject-dependent models.

While a number of related studies have focused on predicting the post-prandial glucose response to different types of meals,

TABLE II. PEARSON CORRELATION AND STATISTICAL SIGNIFICANCE OF THE MACRONUTRIENT ESTIMATES

| | Leave One Subject Out | | | Leave One Meal Out | | |
|---|---|---|---|---|---|---|
| | CHO | Protein | Fat | CHO | Protein | Fat |
| Linear Regression | 0.32** | 0.12 | 0.09 | 0.31** | -0.29 | -0.01 |
| Multitask Neural Network | 0.31** | 0.14 | 0.21 | **0.69*** | **0.23*** | **0.48*** |
| ***: p < 0.001, **: 0.001 ≤ p < 0.05, *: 0.05 ≤ p < 0.1 | | | | | | |

TABLE III. MEAN RMSRE (AND STANDARD DEVIATION) OF THE MACRONUTRIENT ESTIMATES

| | Leave One Subject Out | | | Leave One Meal Out | | |
|---|---|---|---|---|---|---|
| | CHO | Protein | Fat | CHO | Protein | Fat |
| Linear Regression | 0.46 (0.31) | 0.61 (0.25) | 0.62 (0.30) | 1.51 (1.70) | 4.57 (4.44) | 2.64 (4.57) |
| Multitask Neural Network | 0.45 (0.19) | 0.57 (0.24) | 0.54 (0.21) | **0.39 (0.08)** | **0.54 (0.09)** | **0.51 (0.15)** |

this work predicted meal composition from CGM signals. This contribution enables automated logging of macronutrient composition for participants who wear a CGM, allow ing for more accurate and timely monitoring of patient diet.

ACKNOWLEDGMENT

This work was supported, in part, by funding from the National Science Foundation Engineering Research Center for Precise Advanced Technologies and Health Systems for Underserved Populations (PATHS-UP) (Award #1648451).